\documentclass[aps,pra,reprint]{revtex4-1}
\usepackage{graphicx}
\usepackage{dcolumn}
\usepackage{bm}
\usepackage{hyperref}
\usepackage{amssymb}
\usepackage{amsmath}
\usepackage{epsf}
\usepackage[caption=false]{subfig}
\usepackage{epstopdf}
\usepackage{amsfonts}
\usepackage{color}

\usepackage{pst-all}
\usepackage [english]{babel}
\usepackage [autostyle, english = american]{csquotes}
\MakeOuterQuote{"}

\begin{document}
\title{Stochastic dynamics of a non-Markovian random walk in the presence of resetting}
\author{Upendra Harbola}
\affiliation{Department of Inorganic and Physical Chemistry, Indian Institute of 
Science, Bangalore, 560012, India.}
\date{\today}
\begin{abstract}
The discrete stochastic dynamics of a random walker in the presence of resetting and memory is analyzed.  
Resetting and memory effects may compete for certain parameter regime and lead to 
significant changes in the long time dynamics of the walker. 
Analytic exact results are obtained for a model memory where 
the walker remembers all the past events equally.
In most cases, resetting effects dominate at long times and dictate the asymptotic dynamics. 
We discuss the full phase diagram of the asymptotic dynamics and the resulting changes due to the 
resetting and the memory effects. 
\end{abstract}
\maketitle

\section{Introduction}

Stochastic dynamics finds applications in a wide variety of physical systems starting from molecular length scales to interstellar distances\cite{RMP85-135-2013,Chandrasekhar}. 
In the simplest case, the stochastic dynamics is diffusive as characterized by linear increase in time of the variance of displacement. In many natural and man-made systems however
this simple picture breaks down and a nonlinear growth in time is observed, which is termed as sub-diffusive or super-diffusive, that is, slower or faster than the diffusive dynamics,
respectively. There may be several reasons that can give rise to such non-diffusive behaviors but they all can be put into two categories: spatial constraints and temporal or memory 
effects.  Several stochastic models have been proposed  to account for these spatial and temporal effects \cite{gen1,gen2,gen3,gen4,gen5}. 

An important property of a  stochastic process is the first-passage-time distribution (FPTD) \cite{fptd, PhysicaA390-1841-2011}. 
This is the distribution of times that the stochastic 
trajectories take to reach a certain point for the first time during their time evolution.  This has important implications in many 
areas of physics, chemistry, ecology, and finance \cite{Redner, Metzler}. A modified stochastic process that involves resetting positions 
at random times to the initial position has been proposed \cite{MajumdarPRL2011} and analyzed \cite{SabhapanditPRE2015, TopicalReview,Jayanawar}.
This so-called stochastic resetting model is applicable to many natural and man-made systems that involve random hopping of variable lengths 
such as in facilitated diffusion of a protein on DNA in search of a target sequence  \cite{NucAcidRes-Marko2004, BrayAdvPhys2013}, 
enzymatic activity\cite{RotartPRE2015}, and has been realized  in experiments \cite{RoichmanPCL2020, BesgaPRR2020}. 
It is known that random resetting of position leads to significant qualitative changes in the FPTD \cite{MajumdarPRL2011}, giving rise to finite 
moments of FPTD, which are otherwise not defined for the simplest (Markovian) diffusive motion.  

In this work, we consider a non-Markovian discrete stochastic dynamics in the presence of random resetting to arbitrary positions. At any instant of time, 
the forward and backward steps of unit length are determined based on the memory of the past such events. We first derive a Fokker-Planck (FP) equation for 
this walk, without resetting, which leads to a Gaussian probability distribution with mean and variance both displaying power-law increase in time. This 
results in a FPTD with no finite moments and  asymptotically decays as $t^{-3/2}$ in the diffusive regime and $t^{-1-\gamma}$ ($1/2<\gamma<1$) in the 
super-diffusive regime. Introducing the resetting process in this non-Markovian dynamics leads to substantial changes in the dynamics: both the mean position and 
the variance may show exponential increase in time. Different regimes 
emerge where the memory and the resetting processes tend to oppose or support each other. We find that in most cases, at long times the resetting process dominates 
and the asymptotic dynamics is solely governed by the resetting effects.  A non-trivial analytic solution for the probability distribution is obtained by solving the 
corresponding FP equation.

\section{Model and results}
Consider a simple random walker which can take jumps of unit length to its right ($+1$)  and left ($-1$) sides with probability 
$p$ and $1-p$, respectively, on a one-dimensional space. This is a Markov random walk that leads to a diffusive dynamics where, 
at long times, mean and variance of the walker's position vary linearly in time \cite{RiskenBook}.

\subsection{Introduction of memory}

Now let us introduce a particular type of memory in this walk. We consider a case where walker remembers all its past steps with equal probability. At any instant 
of time, it picks-up a past remembered event and performs the same with probability $p$ and opposes it with probability $q$, $p+q=1$.  The probability that the 
$n$th step will be $\sigma_n=+1$ or $-1$ now depends on the entire history of previous jumps $\{\sigma_{n-1}\}=(\sigma_1,\sigma_2,\cdots,\sigma_{n-1})$.
\begin{eqnarray}
\label{eq-1}
P[\sigma_{n+1} =\sigma|\{\sigma_n\}] &=& \frac{1}{2n}\sum_{k=1}^n \left(1+\sigma\gamma\sigma_k\right) \nonumber\\
&=& \frac{1}{2} +\frac{\sigma\gamma}{2n}x_n
\end{eqnarray}
where $\sigma=\pm 1$, $\gamma=p-q$, and $x_n=\sum_{k=1}^n\sigma_k$ is the position after $n$ steps. 
For simplicity, we consider the first step  $\sigma_1=+1$ with unit probability.
Similar random walk models have been studied in Refs. \cite{UH-1}
and \cite{UH-2} which include the possibility of $\sigma=0$ as well. Here we consider a simpler walk without $\sigma=0$ as our motivation is to 
study the effect of memory on simple resetting walk.    

From Eq. (\ref{eq-1}), we find $\langle\sigma_n\rangle = \gamma x_{n-1}/(n-1)$, which upon using in $x_n=\sigma_n+x_{n-1}$ gives a recursion relation 
for the mean position that gives,
\begin{eqnarray}
\label{eq-2}
\langle x_n\rangle &=& \left(1+\frac{\gamma}{n-1}\right) \langle x_{n-1}\rangle \nonumber\\
&=& \frac{\Gamma(n+\gamma)}{\Gamma(n)\Gamma(1+\gamma)}.
\end{eqnarray} 
Thus, unlike the simple Markovian case, the mean position varies nonlinearly in time (here time is same as $n$) and for long time, 
$\langle x_n\rangle \sim n^\gamma/\Gamma(1+\gamma)$.

In order to compute average square position, we note that $\langle \sigma_n^2\rangle =1 \forall n$. This gives,
\begin{eqnarray}
\label{eq-3}
\langle x_n^2\rangle &=& 1+\frac{n+2\gamma-1}{n-1}\langle x_{n-1}^2\rangle\nonumber\\
 &=& \frac{2\gamma}{(2\gamma-1)\Gamma(n)}\left[\frac{\Gamma(n+2\gamma)}{\Gamma(1+2\gamma)}-\frac{\Gamma(n+1)}{2\gamma}\right].
\end{eqnarray}
For asymptotic times, 
\begin{eqnarray}
\label{eq-4}
\langle x_n^2\rangle \sim \frac{1}{2\gamma-1}\left(\frac{n^{2\gamma}}{\Gamma(2\gamma)}-n\right).
\end{eqnarray} 
Clearly, the variance $\Delta x_n^2 =\langle x_n^2\rangle-(\langle x_n\rangle)^2$ at long times is
\begin{eqnarray}
\label{eq-5}
\Delta x_n^2 \sim \left(\frac{1}{(2\gamma-1)\Gamma(2\gamma)}-\frac{1}{\Gamma^2(1+\gamma)}\right) n^{2\gamma} -\frac{n}{2\gamma-1},
\end{eqnarray} 
which varies nonlinearly in time if $\gamma >1/2$.

From Eq. (\ref{eq-5}), we therefore conclude that, due to the memory, the walker's asymptotic dynamics shows a phase transition from diffusive to super-diffusive 
as $\gamma$ is increased beyond $1/2$. In the super-diffusive regime, the variance grows in time with a power-law with exponent $2\gamma$.
Indeed, as $\gamma\to 1$, variance vanishes as the particle motion becomes deterministic, $\langle x_n\rangle =n$ and 
$\langle x_n^2\rangle = n^2$. 

The probability $P(x,t)$ to find the walker at position $x_n\equiv x$ after "time" $t\equiv n$ satisfies the following rate equation (see Appendix)
\begin{eqnarray}
\label{eq-6}
P(x,t) &=& \frac{1}{2}\left(1+\frac{\gamma}{t-1}(x-1)\right)P(x-1,t-1) \nonumber\\
&+& \frac{1}{2} \left(1-\frac{\gamma}{t-1}(x+1)\right)P(x+1,t-1).
\end{eqnarray}
Note that for a given time $t$, $-t+2\leq x \leq t$ and 
$P(x,t)$ satisfies the initial condition $P(x,1)=\delta(x-1)$ together with the boundary conditions 
\begin{eqnarray}
\label{eq-6a}
P(t,t) &=& \left(\frac{1+\gamma}{2}\right)^{t-1}\nonumber\\ 
P(-t+2,t) &=&\frac{1-\gamma}{1+\gamma}\left(\frac{1+\gamma}{2}\right)^{t-1} \frac{\Gamma(t-\frac{2\gamma}{1+\gamma})}{\Gamma(t)\Gamma(\frac{2}{1+\gamma})}.
\end{eqnarray}

The rate Eq. (\ref{eq-6}) leads to a Fokker-Planck equation (see Appendix), 
\begin{eqnarray}
\label{eq-7}
\frac{\partial}{\partial t}P(x,t) = - \frac{\partial}{\partial x}\left[\frac{\gamma x}{t} P(x,t)\right]
+\frac{1}{2} \frac{\partial^2}{\partial x^2} P(x,t),
\end{eqnarray}
for large $t$. This FP equation is same as of a Brownian particle moving in a time-dependent harmonic potential $V(x,t)=(\gamma/t) x^2$. 
In the diffusion regime ($\gamma<1/2$), the mean position of the particle at time $t$ is much smaller compared to the boundary points which are far and can be approximately  
considered at $x=\pm \infty$. In this case,  the FP equation has a well known \cite{PRE61-R4675-2000, EJP-37-065101-2016} Gaussian solution with the mean 
and the variance changing in time according to Eqs. (\ref{eq-2}) and (\ref{eq-5}) at long times, respectively. 
\begin{eqnarray}
\label{eq-7a}
P(x,t) = \sqrt{\frac{2\gamma-1}{2\pi(t^{2\gamma}-t)}} e^{-\frac{(1-2\gamma)(x-t^\gamma)^2}{2(t-t^{2\gamma})}}
\end{eqnarray}
with the initial condition $P(x,t\to1)=\delta(x-1)$ . 

In the super-diffusive regime $\gamma>1/2$, the mean position may lie closer to the boundary points 
(for $\gamma=1$, the dynamics is ballistic) which must be taken into account while solving the FP equation. 
In this case the position distribution may deviate significantly from the Gaussian
function and Eq. (\ref{eq-7a}) may no longer be a valid approximate solution. 
The boundary conditions needed in solving the FP equation include values of $P(x,t)$ at $x=t$ and $x=-t+2$, already given in Eq. (\ref{eq-6a}), and also the derivatives
$\partial P(x,t)/\partial x\approx (P(x,t)-P(x-2,t))/2$ defined at the boundary points. These derivatives in the long time limit can be 
approximated as (see Appendix)
\begin{eqnarray}
\label{eq-6b}
\left.\frac{\partial P(x,t)}{\partial t}\right|_{x=t} &\approx&  \frac{1}{2}\left(\frac{1+\gamma}{2}\right)^{t-1}\left[1-t\left(\frac{1-\gamma}{1+3\gamma}\right)\right]\nonumber\\
\left.\frac{\partial P(x,t)}{\partial t}\right|_{x=-t+2} &\approx& \frac{t^{\frac{1-\gamma}{1+\gamma}}}{2}
\frac{(1-\gamma)^2}{(1+\gamma)(1+3\gamma)}  \left(\frac{1+\gamma}{2}\right)^{t-1}.
\end{eqnarray}
Note that for $\gamma\to 1$, the derivative at $x=t$ approaches to a constant value $1/2$ while the derivative at $x=-t+2$ vanishes.
\begin{figure}[h]
\includegraphics[width=8cm]{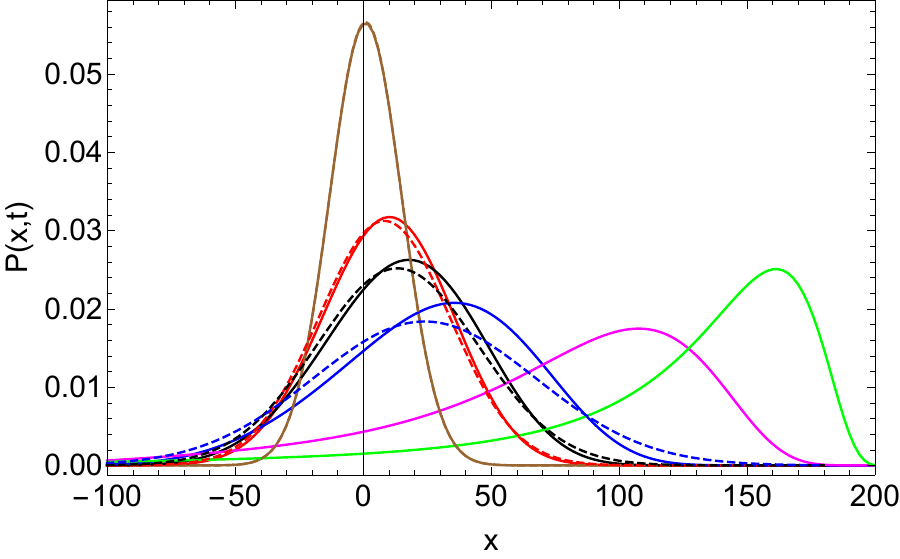}
\caption{Exact probability distribution (solid) obtained by numerical solution of Eq. (\ref{eq-6}) for $\gamma=0.0$ (brown),  $\gamma=0.4$ (red), $\gamma=0.5$ (black) $\gamma=0.6$ (blue), $\gamma=0.8$ (magenta), and $\gamma=0.9$ (green) at $t=200$. Dashed curves show Gaussian approximate solution for $\gamma=0.0, 0.4, 0.5, 0.6$.
For $\gamma=0.0$, the two results agree perfectly.}
\label{figpdf}
\end{figure}

Equation (\ref{eq-7}) can be solved numerically using iterative method. In Fig. (\ref{figpdf}), we depict the numerical solutions for different  memory ($\gamma$) 
values, and the corresponding approximate Gaussian  solution of FP equation are also shown for comparison for low memory. It is clear that in the diffusive regime 
$\gamma<1/2$, the approximate result seems to be working all right while in the super-diffusive regime, the approximate Gaussian result completely breaks down as the 
exact result is highly non-Gaussian and shows a long tail behavior.

An interesting property related to a random walk is the first passage time (FPT): the time that the walker takes to reach a certain point first time. This has wide 
applications such as in drug delivery, spontaneous chemical reaction rates, etc. The probability $F(x,t)$ to arrive first time a point at a distance $x$ from 
the initial position in time  $t$ is defined as $F(x,t)=-\frac{d}{d t} S(t|x)$ where survival probability $S(t|x)$ having an absorbing boundary at $x$ is obtained 
within the image method \cite{PhysicaA390-1841-2011} (see Appendix) by placing the absorbing boundary at $x=0$ and shifting the initial position of the walker at $x$. 
 Note that in presence of time-dependent drift, the FPT distribution depends on whether the drift is towards or away from the absorbing boundary (or point). 
 For the case $0<\gamma<1/2$,  the FPT distribution is obtained as, 
\begin{eqnarray}
\label{eq-fpt-1a}
F(x,t) &=& 2[x(1-2\gamma t^{2\gamma-1})\pm t^\gamma(1-2\gamma)] \nonumber\\
&\times&\sqrt{\frac{(2\gamma-1)}{2\pi(t^{2\gamma}-t)^3}} e^{-\frac{(2\gamma-1)}{t^{2\gamma}-t}\frac{(x\pm t^\gamma)^2}{2}}
\end{eqnarray}
where the upper (lower) sign is for the case when the drift is away from (towards) the absorbing boundary. 
$F(x,t)$ decays asymptotically as $~t^{-(3/2-\gamma)}$ in the diffusive regime and suppresses quickly (with Gaussian weight) 
with increasing $x$, the distance between the initial position and the absorbing point, indicating that 
positions sufficiently far from the initial point and away from the direction of the drift, are reached with vanishing probabilities.   
When $\gamma=0$, the FPT reduces to that for an unbiased random walk. Note that even for $\gamma=0$, the $\pm$ sign still remain because, unlike the standard case,  
the initial position is at $x=+1$, that is, position of the absorbing point, which is at a distance $x$ from the initial position,  differ by unity depending on whether it is located 
on the right-side or the left-side of the initial position. 

For $(\gamma < 0)$, the drift is always towards $x=0$, starting from $x=1$. The FPT is given by Eq. (\ref{eq-fpt-1a}) 
and decays asymptotically as, $~ t^{-3/2}$.  This case is similar to the standard diffusive process.
Thus, memory introduces qualitative changes in  the dynamics (diffusion to super-diffusion). However, these changes are not reflected in the FPTD, which 
like the standard diffusive case,  does not posses any finite moments.

\begin{figure}[h]
\includegraphics[width=8cm]{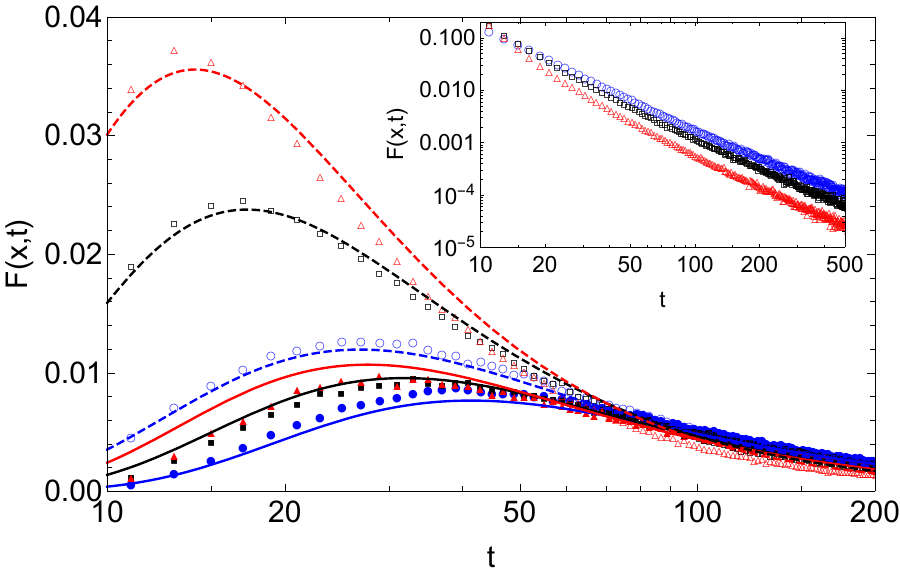}
\caption{FPT distribution with absorbing boundary at $x=10$ in the diffusive regime  for $\gamma=0$(blue), 
$\gamma=0.2$ (black), and $\gamma=0.3$ (red).  Filled (empty) points denote simulation results when the drift is away from 
(towards) the absorbing point. Solid (dashed) curves are the  analytic result from Eq. (\ref{eq-fpt-1a}). Inset shows simulation 
results in the super-diffusive regime for $\gamma=0.6$ (blue), $0.7$ (black), and $0.8$ (red).} 
\label{fig5}
\end{figure}

Figure (\ref{fig5}) shows FPT distributions for various values of the memory ($\gamma$) in the diffusive regime ($\gamma<1/2$). 
We note that, in general,  the peak of the distribution shifts towards the smaller times with increasing memory. That is,  it becomes more probable
to reach the absorbing point at earlier times. This sounds counter-intuitive for the case when the drift, which increases with $\gamma$, is away from the 
absorbing point. However,  fluctuations (variance) also grow with $\gamma$ and help to reach the boundary point faster at smaller times. 
Of course, for asymptotically large times,  the variance grows linearly in time as we are in the diffusive regime. Comparatively, when the drift is towards the 
absorbing point, the optimal time to reach the boundary decreases more significantly and the probability increases more as the memory in increased. 
In this case, both fluctuations and drift help reach the absorbing boundary earlier as the memory is increased.
In the super-diffusive regime (inset), the FPT distribution shows almost an exponential decay in time. 

\subsection{Introduction of resetting}

We now consider the case where we are allowed to reset the position at random times to a new position which depends on the present position as discussed below. 
Thus at any point in time ($n$), we have three
possibilities of $\sigma_n = +1, -1$, and a resetting jump which is fraction $f$ of its current position, $\sigma_n=-f x_{n-1}$. We would like to understand how resetting
and memory effects interplay and control the dynamics.  Note that for the standard resetting walk we need to put $f=1$. 
A Markov resetting walk for all $f$ values was analyzed in Ref. \cite{UH-3} and with partial resetting in \cite{RoichmanPRE2022}. 
Singh $et.$ $al.$ \cite{SinghPRE2022} have analyzed resetting dynamics in terms of moments of resetting time-distribution.

We assume that the walker remembers only $\pm 1$ steps and does not remember resetting step, which is considered a Markovian process with rate $r$. Thus at any time
we have $p+q+r=1$.

In the presence of resetting, Eq. (\ref{eq-1}) modifies to
\begin{eqnarray}
\label{eq-8}
P[\sigma_{n+1}=\sigma|\{\sigma_n\}] = \frac{1-r}{2} +\frac{\sigma\gamma}{2n}x_n.
\end{eqnarray}
Following steps that lead to Eq. (\ref{eq-2}), we now have 
\begin{eqnarray}
\label{eq-9}
\langle x_n\rangle = \frac{(1-fr)^n}{\gamma \Gamma(n)}\frac{\Gamma(n+\frac{\gamma}{1-rf})}{\Gamma(\frac{\gamma}{1-rf})}.
\end{eqnarray}
In the asymptotic limit of $n$, $\langle x_n\rangle \sim (1-fr)^n n^{\gamma/(1-fr)}$ where the first term $(1-fr)^n$ is due to the resetting while the second term
arises due to the memory $(\gamma)$. 
For $0<fr<1$, there is a competition between the resetting and the memory effects. The former tries to bring the walker closer to the initial position while the latter  
tries to move it away. The resetting takes over the memory and the mean position of the walker approaches to $x=0$ at asymptotic times. 
Note that the resetting also modifies the memory effects by rescaling the parameter $\gamma\to\gamma/(1-fr)$.
For $1<fr<2$ and $\gamma>0$, both resetting and memory work in tendem to bring the average position of the walker to the initial position.  
For $fr=2$, the average position of the walker oscillates around its initial value with amplitude slowly decreasing (increasing) with power-law weight $n^{-\gamma}$
for $\gamma>0 (\gamma<0)$.

For $fr >2$ and $\gamma<0$, both, resetting and memory, drag the walker away from the initial position while for $\gamma>0$, memory tries to keep the 
walker's position close to the initial value. In both the cases ($\gamma>0$ and $\gamma<0$), the resetting process dominates over the memory and the walker drifts away from 
its initial position at an exponential rate. 

\begin{figure}[h]
\includegraphics[width=8cm]{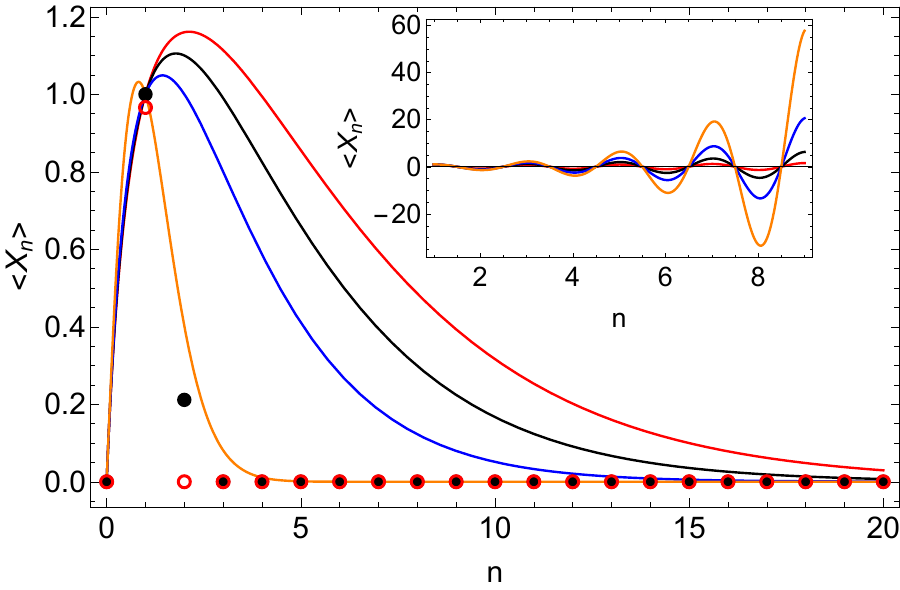}
\caption{Mean position of the walker with time for $\gamma=0.4, r=0.2$ and $f=1.2,1.5, 2.0, 5.0, 6.0, 7.0$ from top (red) to bottom (red circle).
Results for $f=6.0, 7.0$ are shown by dots as the mean position is sharply peaked at $n=1$.
Inset shows exponential growth in the mean position for $fr >2$, $f=11$ (red),$12$(black),$13$(blue),$14$(orange). }
\label{fig1a}
\end{figure}

Time evolution of the mean position of the walker, Eq. (\ref{eq-9}), for various resetting scenarios discussed above is depicted in Fig. (\ref{fig1a}).

For the second moment, $\langle x_n^2\rangle$, we obtain a recurrence relation $\langle x_{n}^2\rangle=1-r+(a+2\gamma/(n-1))\langle x_{n-1}^2\rangle$
where $a = 1+rf^2-2rf$. The recurrence relation can be solved iteratively to obtain,
\begin{eqnarray}
\label{eq-10}
\langle x_n^2\rangle &=&  \frac{a^{n-1}\Gamma(n+\frac{2\gamma}{a})}{\Gamma(n)\Gamma(1+\frac{2\gamma}{a})}
\left[1+\frac{1-r}{a+2\gamma}F\left(1, 2, 2+\frac{2\gamma}{a},\frac{1}{a}\right)\right]\nonumber\\
&-&
(1-r)\left[F\left(1,n,n+\frac{2\gamma}{a},\frac{1}{a}\right)-1\right]
\end{eqnarray}
where $F(b,c,d,e)$ is the Hypergeometric function. Parameter $a>0$ controls the resetting effect. 

Similar to the case of mean position discussed above, resetting modifies the fluctuation by rescaling the memory effects 
$\gamma \to \gamma/a$. Thus for $0<f<2$ (for which $0< a <1$), the effective value of $\gamma$ increases due to the resetting process. 
That is, due to resetting, the effective memory tends to drag the walker more into the super-diffusive regime by increasing the value of $\langle x_n^2\rangle$ in time. 
On the other hand, the explicit dependence on the resetting which is contained in the factor $a^{n-1}$ tends to exponentially suppress the value of the fluctuation. 
The result is that, in the long time,  the fluctuation approaches a time-independent value determined by the second term on the right hand side of Eq. (\ref{eq-10}).
For $f>2 (a>1)$,  the effective $\gamma$ value decreases, that is, increase in the fluctuation with time due to the memory is suppressed while the explicit resetting
dependence tends to exponentially increase the fluctuation in time. Thus the "effective memory" and the resetting both compete against each other with the result that 
in the long time, the fluctuation and also the variance increase exponentially and the walker is in the super-diffusive regime. 
\begin{figure}[h]
\label{fig1}
\includegraphics[width=8cm]{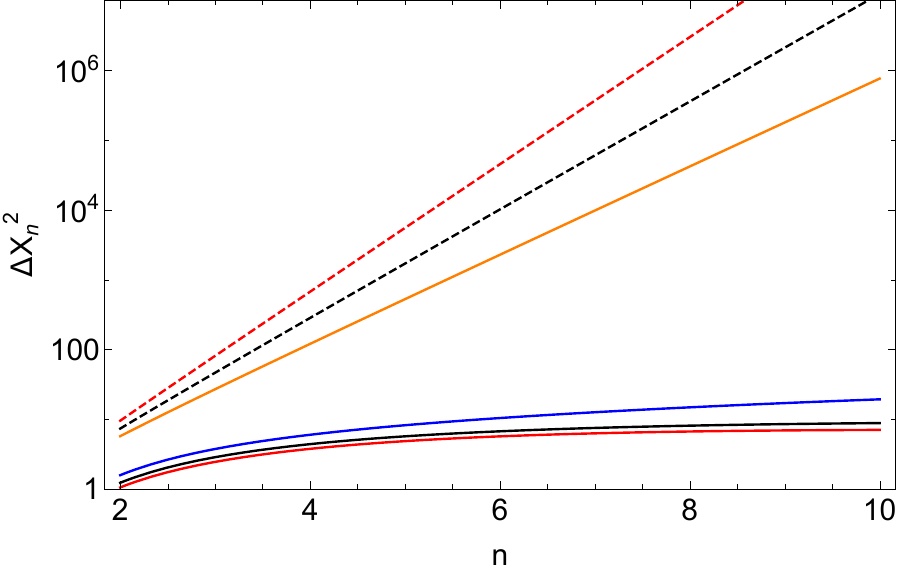}
\caption{Variance in the walker's position for $\gamma=0.4, r=0.2$ and $f=1.2,1.5, 2.0$ (solid red, black and blue curves) in the diffusive regime ($fr<2$) and 
$f= 5.0, 6.0, 7.0$ (orange, dashed black and  dashed red curves) in the super-diffusive regime.}
\end{figure}

Interestingly, for $f=2~ (a=1)$, 
although resetting does affect the mean position of the walker as discussed above,  it does not however influence the long time dynamics 
of  the fluctuation $\langle x_n^2\rangle$, which is determined only by the memory effects according to the 
discussion in the previous section. This can be understood as follows.
For $f=2$, the long jumps $\sigma=-f x_n$ do not change the distance of the particle from its initial position
and only take it from one-side to the other side of the initial position, affecting the mean position but not the fluctuation.  

In the presence of resetting, the rate equation, Eq. (\ref{eq-6}), modifies to 
\begin{eqnarray}
\label{eq-11}
P(x,t) &=& \frac{1}{2}\left(1-r+\frac{\gamma}{t-1}(x-1)\right)P(x-1,t-1) \nonumber\\
&+& \frac{1}{2} \left(1-r-\frac{\gamma}{t-1}(x+1)\right)P(x+1,t-1) \nonumber\\
&+& rP(x/(1-f),t-1).
\end{eqnarray}
For  the standard resetting walk, ($f=1$), the rate equation leads to the following Fokker-Planck equation for $t>\mbox{max}(\frac{\gamma}{r},\frac{\gamma}{1-r})$,
\begin{eqnarray}
\label{eq-11a}
\frac{\partial P(x,t)}{\partial t} &=& -\frac{r}{1-r}P(x,t)-\frac{\gamma x}{t(1-r)}\frac{\partial P(x,t)}{\partial x} \nonumber\\
&+&\frac{1}{2}\frac{\partial^2P(x,t)}{\partial x^2}
+ \frac{r}{1-r}\delta(x)
\end{eqnarray}
which can be solved to obtain the following probability distribution.
\begin{eqnarray}
\label{eq-12}
P(x,t) &=& \sqrt{\frac{1-2b}{2\pi}}\left[ \frac{e^{-\frac{r}{1-r}(t-1)}}{\sqrt{t-t^{2b}}}  e^{-\frac{1-2b}{t-t^{2b}}\frac{(x-t^b)^2}{2}}\right.\nonumber\\
&+&\left.\frac{r }{1-r} \int_1^t ds \frac{e^{-\frac{r(t-s)}{1-r}}}{\sqrt{t-t^{2b}s^{1-2b}}} e^{-\frac{1-2b}{t-t^{2b}s^{1-2b}}\frac{x^2}{2}}\right]\nonumber\\
\end{eqnarray}
where $b=\gamma/(1-r)$. For no memory $(\gamma=0)$, Eq. (\ref{eq-12}) reduces to the well known result for the standard resetting 
walk which reaches the steady-state $P(x)\propto e^{-|x|\sqrt{2r/(1-r)}}$. 
The same steady-state is reached even in the presence of the memory since, for $f=1$, the resetting effects dominate at large times. 
It is clear from Eq. (\ref{eq-12}) that the resetting process suppresses the memory effect, the first term in Eq. (\ref{eq-12}),
 exponentially in time, and the second term determines the steady state. 

\begin{figure}[h]
\includegraphics[width=8cm]{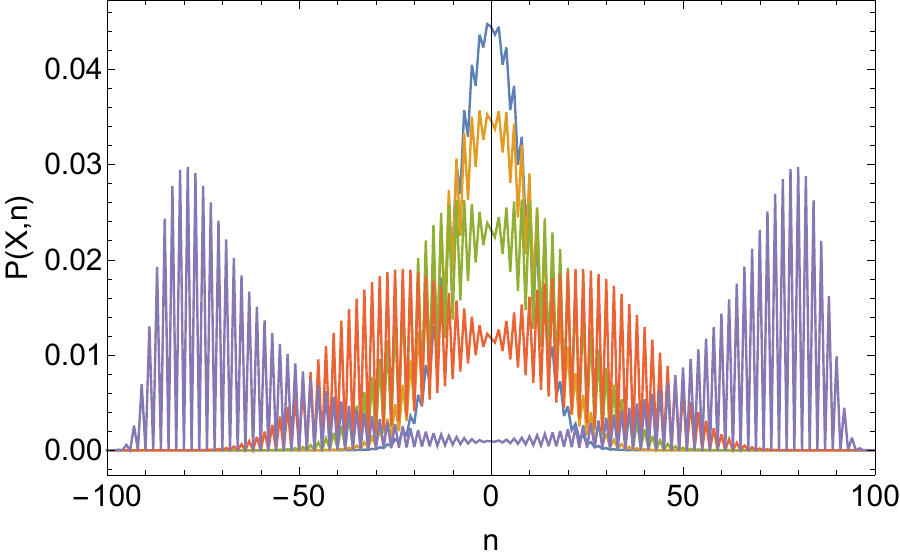}
\caption{Probability distribution of the walker for $f=2$ at time $t=100$ and $r=0.2$ for different memory $\gamma=0.0$ (blue), $0.2$ (orange), $0.4$ (green), $0.6$(red), 
and $0.9$ (purple).}
\label{fig4}
\end{figure}

To explore memory effects for larger values of $f>1$, we numerically solve the rate equation to obtain position distribution. 
In Fig. (\ref{fig4}), we show some results for $f=2$ for different values of the memory. It is clear that in this case, the memory effects are 
significant: for small memory, the walker is most likely to be found at the initial position but as the memory increases, its position distribution develops 
a double-peak structure with most probable positions lying equidistant on either side of the initial position. Average position of the walker is always zero, although
the probability for walker to be at the initial position vanishes as the memory is increased (large $\gamma$).  

\section{Conclusion}
A simple random walk model where particle step at time $t$ is determined by the past events (steps forward and backward) that the walker 
remembers with equal probability is coupled with resetting process, leads to rich dynamics at long times. The memory effects alone give rise to simple diffusive or super-diffusive 
dynamics with no steady-state, while the resetting process, in addition to diffusive and super-diffusive behavior, may also lead to a steady-state at long times. 
The resetting effects are robust, that is, in most cases when there is a competition between the two processes (resetting and memory), the effect of resetting dominates
and dictates the final dynamics.  In the case where resetting involves jumps of twice the current position leading to a swapped position $x\to -x$ and vice versa, 
the fluctuation dynamics remains unaffected by the resetting process while the mean position is strongly affected by the resetting causing the walker to remain localized at its initial position at long times.   In the transient dynamics, memory effects tend to suppress fluctuations.

\section*{Acknowledgements}

Financial support from SERB, India under the Grant No. CRG/2020/0011100 is acknowledged.

\section*{Appendix}

\begin{center}
{\bf Derivation of the rate equation, Eq. (\ref{eq-6})}
\end{center}

In order to derive the rate Eq. (\ref{eq-6}), we define characteristic function 
\begin{eqnarray}
\label{app-1}
Q(\lambda,t) = \langle e^{i\lambda X_t}\rangle =\sum_{X_t} e^{i\lambda X_t} P(x, t)
\end{eqnarray}
 where $P(x,t)$ is the probability to be at $X_t=x$ at time $t$. 
 Using $X_t=X_{t-1}+\sigma_t$ and averaging over $\sigma_t$ for a given history, as defined in Eq. (\ref{eq-8}), we get
\begin{eqnarray}
\label{app-2}
\langle e^{i\lambda X_t}|\{\sigma_{t-1}\} \rangle= e^{i\lambda X_{t-1}} \sum_{n=0}^\infty \frac{(i\lambda)^n}{n!}\langle \sigma_t^n|\{\sigma_{t-1}\}\rangle 
\end{eqnarray}
where $\langle\cdots\rangle_{\sigma_t}$ is used to denote that the averaging is only over the last step $\sigma_t$. 

Using Eq. (\ref{eq-1}), we have
\begin{eqnarray}
\label{app-2a}
\langle \sigma_t^n|\{\sigma_{t-1}\}\rangle&=& \frac{1}{2}(1+(-1)^n)+\frac{\gamma x_{t-1}}{2(t-1)}(1-(-10)^n).
\end{eqnarray}
Substituting this in Eq. (\ref{app-2}), and averaging over all $X_{t-1}$, we get
\begin{eqnarray}
\label{app-3}
Q(\lambda,t) =  \mbox{cos}(\lambda) Q(\lambda,t-1)+ \frac{\gamma}{t-1}\mbox{sin}(\lambda)\frac{\partial}{\partial\lambda} Q(\lambda,t-1).\nonumber\\
\end{eqnarray}

Inverse transforming, $P(x,t)=\int \frac{d\lambda}{2\pi}e^{-i\lambda X_t}Q(\lambda,t)$, and using
\begin{widetext}
\begin{eqnarray}
\label{app-5}
\int \frac{d\lambda}{2\pi}e^{-i\lambda X_t}Q(\lambda(1-f),t-1) &=&  \int \frac{d\lambda}{2\pi}e^{-i\lambda X_t} \sum_{X_{t-1}}e^{i\lambda(1-f)X_{t-1}}P(X_{t-1})\nonumber\\
&=& \sum_{X_{t-1}} P(X_{t-1})\delta_{X_t, (1-f)X_{t-1}}=P(X_t/(1-f)) \equiv P(x/(1-f), t-1)
\end{eqnarray}
\end{widetext}
leads to Eq. (\ref{eq-6}).

\subsection{Derivation of the Fokker-Planck Eq. (\ref{eq-7}) and its solution, Eq. (\ref{eq-7a})}

In the rate Eq. (\ref{eq-6}) we use Taylor expansion to expand $P(x\pm 1,t-1)$ around $P(\pm x, t)$, we have 
$P(x\pm 1,t-1)\approx P(x,t)\pm \frac{\partial P(x,t)}{\partial x}+\frac{1}{2}\frac{\partial^2 P(x,t)}{\partial^2 x}-\frac{\partial P(x,t)}{\partial t}$.
Substituting this in Eq. (\ref{eq-6}) and considering $x,t \gg1$, we obtain the FP Eq. (\ref{eq-7}). 

In order to solve the FP equation for natural boundary conditions, we first Fourier transform to $k$-space, $P(k,t)=\int_{-\infty}^\infty dx e^{ikx}P(x,t)$. This gives
\begin{eqnarray}
\label{app-6}
\frac{\partial P(k,t)}{\partial t} = \frac{\gamma k}{t}\frac{\partial P(k,t)}{\partial k}-\frac{k^2}{2} P(k,t)
\end{eqnarray} 
with the initial condition $P(k,t=1)=e^{ik}$. We next parameterize both $k$ and $t$ by $s$ such that $P(s)\equiv P(k(s),t(s))$, where $1\leq s\leq \infty$ and 
$t$ varies from $1$ to $t$ as $s$ takes values from $s=1$ to $s$. Then
\begin{eqnarray}
\label{app-7}
\frac{dP(s)}{ds} = \frac{\partial P(s)}{\partial t}\frac{\partial t}{\partial s} + \frac{\partial P(s)}{\partial k}\frac{\partial k}{\partial s}.
\end{eqnarray}
Considering $\partial t/\partial s=1$ and $\partial k/\partial s=-\gamma k/t$, so that $t=s$ and $k(s) = k_0s^{-\gamma}$ for constant $k_0$.
These can be trivially inverted to obtain $s=t$ and $k_0=k t^{\gamma}$.
Equations (\ref{app-6}) and (\ref{app-7}) then allow us to write, 
\begin{eqnarray}
\label{app-8}
\frac{dP(s)}{ds} = -\frac{k^2(s)}{2} P(s)  = -\frac{k_0^2 s^{-2\gamma}}{2} P(s),
\end{eqnarray} 
which is solved to obtain $P(s)=P(1)e^{-\frac{k_0^2}{2}\frac{s^{1-2\gamma}-1}{1-2\gamma}}$ with $P(1)=e^{ik_0}$.
Upon transforming back from $s$ to $k,t$ and substituting for $k_0$, we obtain,
\begin{eqnarray}
\label{app-9}
P(k,t) = e^{ikt^\gamma} e^{-\frac{k^2}{2}\frac{t^{2\gamma}-t}{2\gamma-1}}.
\end{eqnarray}
Inverse Fourier transforming Eq. (\ref{app-9}), we obtain the desired result given in Eq. (\ref{eq-7a}).

Next we solve the FP equation with the boundary conditions given in Eqs. (\ref{eq-6a}) and (\ref{eq-6b}).
We again define the Fourier transform $P(k,t)=\int_{-t+2}^t dx e^{ikx}P(x,t)$. Note the finite range $-t+2\leq x\leq t$ 
for a given $t$. This gives,
\begin{eqnarray}
\label{app-10}
\int_{-t+2}^t dx e^{ikx} x\frac{\partial P(x,t)}{\partial x} &=&(t-2)e^{-ik(t-2)}P(-t+2,t)\nonumber\\
& +&  te^{ikt} P(t,t) -\frac{\partial}{\partial k}\left(kP(k,t)\right),\nonumber\\
&\approx& te^{ikt}P(t,t)+te^{-ikt}P(-t+2,t)\nonumber\\
&-& \frac{\partial}{\partial k} k P(k,t)
\end{eqnarray} 
where the last line is for large $t\gg1$ limit.

Similarly, 
\begin{eqnarray}
\label{app-11}
\int_{-t+2}^t dx e^{ikx} \frac{\partial^2 P(x,t)}{\partial x^2} & \approx& - e^{-ikt}\left.\frac{\partial P(x,t)}{\partial x}\right|_{x=-t+2}\nonumber\\
& -& k^2P(k,t) + e^{ikt}\left.\frac{\partial P(x,t)}{\partial x}\right|_{x=t}\nonumber\\
&+&  ike^{ikt}P(t,t)\nonumber\\
&-&ike^{-ikt}P(-t+2,t).\nonumber\\
\end{eqnarray} 

The FP equation in $k$-space is then obtained as,
\begin{eqnarray}
\label{app-12}
\frac{\partial P(k,t)}{\partial t} &=& \frac{\gamma k}{t}\frac{\partial P(k,t)}{\partial k}-\frac{k^2}{2}P(k,t) \nonumber\\
&+& \left(\frac{1+\gamma}{2}\right)^{t-1}f(k,t),
\end{eqnarray}
where $f(k,t)=A_1(k,t)e^{ikt}+A_2(k, t)e^{-ikt}$ with
\begin{eqnarray}
\label{app-13}
A_1(k,t) &=&  -\gamma-\frac{ik}{2}-\frac{t}{4} \frac{1-\gamma}{(1+3\gamma)} \nonumber\\
A_2(k,t) &=&  -\frac{1-\gamma}{1+\gamma}\left((\gamma-\frac{ik}{2})\frac{t^{-\frac{2\gamma}{1+\gamma}}}{\Gamma(\frac{2}{1+\gamma})}
+\frac{1-\gamma}{4(1+3\gamma)}t^{\frac{1-\gamma}{1+\gamma}}\right)\nonumber\\. 
\end{eqnarray}
Following steps that lead from Eq. (\ref{app-6}) to (\ref{app-9}),  we obtain,
\begin{eqnarray}
\label{app-14}
P(k,t) &=& e^{ikt^\gamma} e^{-\frac{k^2}{2}\frac{t^{2\gamma}-t}{2\gamma-1}} +
\frac{2}{1+\gamma} \int_{1}^t d\tau \left(\frac{1+\gamma}{2}\right)^{\tau}  \nonumber\\
&\times& e^{-\frac{k^2}{2}\frac{t^{2\gamma}\tau^{1-2\gamma}-t}{2\gamma-1}} f(kt^\gamma\tau^{-\gamma},\tau).
\end{eqnarray}
This upon inverse Fourier transforming  to $x$-space gives,
\begin{eqnarray}
\label{app-15}
P(x,t) &=& \sqrt{\frac{2\gamma-1}{2\pi(t^{2\gamma}-t)}} e^{-\frac{2\gamma-1}{t^{2\gamma}-t}\frac{(x-t^\gamma)^2}{2}}\nonumber\\
&+& \frac{2}{1+\gamma} \int_{1}^t d\tau \left(\frac{1+\gamma}{2}\right)^{\tau}\sqrt{\frac{2\gamma-1}{2\pi(t^{2\gamma}\tau^{1-2\gamma}-t)}} \nonumber\\
&\times& \left[{\cal A}_1(\tau) e^{-\frac{(2\gamma-1)(x-t^\gamma\tau^{1-\gamma})^2}{2(t^{2\gamma}\tau^{1-2\gamma}-t)}} \right.\nonumber\\
&+&\left.{\cal A}_2(\tau) e^{-\frac{(2\gamma-1)(x+t^\gamma\tau^{1-\gamma})^2}{2(t^{2\gamma}\tau^{1-2\gamma}-t)}} 
\right]\nonumber\\
\end{eqnarray}
where
\begin{eqnarray}
\label{app-16}
{\cal A}_1 &=& -\gamma-\frac{1-\gamma}{1+3\gamma} \frac{\tau}{4} 
+\frac{t^\gamma\tau^{-\gamma}}{2}\frac{(2\gamma-1)(x-t^\gamma\tau^{1-\gamma})}{t^{2\gamma}\tau^{1-2\gamma}-t}\nonumber\\
{\cal A}_2 &=& -\frac{1-\gamma}{1+\gamma}\left[\gamma\frac{\tau^\frac{-2\gamma}{1+\gamma}}{\Gamma(\frac{2}{1+\gamma})}+\frac{1-\gamma}{4(1+3\gamma)}
\tau^{\frac{1-\gamma}{1+\gamma}}\right.\nonumber\\
& -&\left. \frac{t^\gamma\tau^{-\gamma}}{2}\frac{2\gamma-1}{t^{2\gamma}\tau^{1-2\gamma}-t}(x+t^\gamma\tau^{1-\gamma})
\right]
\end{eqnarray}

\subsection{Derivation for FPT distribution, Eq. (\ref{eq-fpt-1a})}

In order to compute the FPT of the walker, we consider an absorbing boundary placed at a distance $x_0+1$ from the initial position $x=1$. 
We then displace the $x$- coordinate such that the absorbing boundary is at $x=0$. Solution of the FP Eq. (\ref{eq-7})  in the displaced
coordinates with natural boundary conditions is obtained by changing $x\to x-x_0$. 
To find solution with the absorbing boundary at $x=0$, we follow the "image" method. We thus
consider a mirror image of the walker on the  other side of the origin (absorbing point). We first consider the  case when the drift of the dynamics
is away from the absorbing point. In this case, both the real walker and its image drift away from the absorbing point. Thus a solution is given in terms of 
the linear sum of the solutions for the real walker and for the image point. 
\begin{eqnarray}
\label{app-17}
\tilde{P}(x,t) &=& \sqrt{\frac{2\gamma-1}{2\pi(t^{2\gamma}-t)}} \nonumber\\
&\times&\left( e^{-\frac{(2\gamma-1)(x-x_0-t^\gamma)^2}{2(t^{2\gamma}-t)}}
-e^{-\frac{(2\gamma-1)(x+x_0+t^\gamma)^2}{2(t^{2\gamma}-t)}}
\right).
\end{eqnarray}
The survival probability $S(t)$ is obtained by integrating $\tilde{P}(x,t)$ over the region $0<x\leq \infty$.This gives survival
probability in terms of the error function as, $S(x_0, t) =\mbox{erf}\left(\sqrt{\frac{2\gamma-1}{2(t^{2\gamma}-t)}}(x_0+t^\gamma)\right)$, where
$\mbox{erf}(x)=(2/\sqrt{\pi})\int_0^x dy e^{-y^2}$.
The FPT distribution is then given by $F(x,t)=-dS(x, t)/dt$. This yields Eq. (\ref{eq-fpt-1a}).

When the drift is towards the absorbing boundary, we need to change $x_0$ to $-x_0$ in Eq. (\ref{app-17}) so that the average position of the walker (drift) 
starting from $-x_0+1$ moves towards the absorbing point at $x=0$. Following the same steps as for the case when the drift is away from the absorbing boundary, we obtain
FPT distribution as given in Eq. (\ref{eq-fpt-1a}).


\begin{thebibliography}{99}

\bibitem{RMP85-135-2013}
P. C. Bressloff and J. M. Newby, Rev. Mod. Phys. {\bf 85}, 135 (2013).

\bibitem{Chandrasekhar}
S. Chandrasekhar, Rev. Mod. Phys. {\bf 15}, 1 (1943).

\bibitem{gen1}
P. Hanggi, Z. Phys. B 31, 407 (1978).

\bibitem{gen2}
V. S. Volkov and V. N. Pokrovsky, J. Math. Phys. 24, 267 (1983).

\bibitem{gen3}
R. Metzler, E. Barkai, and J. Klafter, Phys. Rev. Lett. 82, 3563 (1999);

\bibitem{gen4}
N. Madras and G. Slade, The Self Avoiding Walk (Birkhauser, Boston, 1993);

\bibitem{gen5}
U. Harbola, N. Kumar, K. Lindenberg, Phys. Rev. E {\bf 90}, 022136 (2014).

\bibitem{fptd}
G. H. Weiss, Adv. Chem. Phys. {\bf 13}, 1-18 (1966). 

\bibitem{PhysicaA390-1841-2011}
A. Molini, P. Talkner, G. G. Katul and A. Porporato, Physica A {\bf 390}, 1841 (2011).

\bibitem{Redner}
S. Redner, {\em A Guide to First Passage Processes}, Cambridge University Press, Cambridge UK, 2001.

\bibitem{Metzler}
R. Metzler, G. Oshanin, and S. Redner (ed), {\em First Passage Phenomena and Their Applications}, Singapore, World Scientific, 2014.

\bibitem{MajumdarPRL2011}
M. R. Evans and S. N. Majumdar,  Phys. Rev. Lett. {\bf 106}, 160601 (2011).

\bibitem{SabhapanditPRE2015}
S. N. Majumdar, S. Sabhapandit, and G. Schehr, Phys. Rev. E {\bf  91}, 052131 (2015). 

\bibitem{TopicalReview}
M. R. Evans, S. N. Majumdar, and G. Schehr, J. Phys. A: Math. Theor. {\bf 53}, 193001 (2020);

\bibitem{Jayanawar}
S. Gupta and A. M. Jayannavar,  Front. Phys. {\bf 10}, 789097 (2022).

\bibitem{NucAcidRes-Marko2004}
S. E. Halford and J. F. Marko, Nucleic Acids Research  {\bf 32}, 3040 (2004).

\bibitem{BrayAdvPhys2013}
A. J. Bray, S. N. Majumdar, and G. Schehr, Adv. Phys. {\bf 62}, 225 (2013).

\bibitem{RotartPRE2015}
S. Reuveni, M. Urbakh, and J. Klafter, Proc Natl.  Acad. Sci U. S. A. {\bf 111}, 4391 (2014). 

\bibitem{RoichmanPCL2020}
O. Tal-Friedman, A. Pal, A. Sekhon, S. Reuveni, and Y. Roichman, J. Phys. Chem. Lett. {\bf 11},  7350 (2020).

\bibitem{BesgaPRR2020}
B. Besga, A. Bovon,  A. Petrosyan, S. N. Majumdar, and S. Ciliberto, Phys. Rev. Res. {\bf  2}, 032029  (2020). 

\bibitem{RiskenBook}
H. Risken, {\em The Fokker-Planck Equation: Methods of Solution and Applications}, Springer; 2nd ed. (1996).

\bibitem{UH-1}
N. Kumar, U. Harbola, and K. Lindenberg, Phys. Rev. E {\bf 82}, 021101 (2010).

\bibitem{UH-2}
U. Harbola, N. Kumar, and K. Lindenberg, Phys. Rev. E {\bf 90}, 022136 (2014).

\bibitem{PRE61-R4675-2000}
F. Lillo and R. N. Mantegna Phys. Rev. E {\bf 61}, R4675 (R) (2000).

\bibitem{EJP-37-065101-2016}
K. S. Fa, Eur. J. Phys. {\bf 37}, 065101 (2016).

\bibitem{UH-3}
U. Harbola, Physical Review E {\bf 108} (1), 014135 (2023).

\bibitem{RoichmanPRE2022}
O. Tal-Friedman, Y. Roichman, and S. Reuveni, Phys. Rev. E {\bf  106}, 054116 (2022).


\bibitem{SinghPRE2022}
R. K. Singh, K. G\'{o}reska, and T. Sandev, Phys. Rev. E {\bf 105}, 064133 (2022).






\end{thebibliography}
\end{document}